\documentclass{article}
\usepackage{amsmath, amssymb, amsfonts}
\title{Anisotropic fluid inside a relativistic star }
\author{Hristu Culetu, \\Ovidius University, Dept.of Physics, \\B-dul Mamaia 124, 900527 Constanta, Romania, \\e-mail : hculetu@yahoo.com}

\begin{document}
\numberwithin{equation}{section}
\pagenumbering{arabic}
\maketitle
\newcommand{\fv}{\boldsymbol{f}}
\newcommand{\tv}{\boldsymbol{t}}
\newcommand{\gv}{\boldsymbol{g}}
\newcommand{\OV}{\boldsymbol{O}}
\newcommand{\wv}{\boldsymbol{w}}
\newcommand{\WV}{\boldsymbol{W}}
\newcommand{\NV}{\boldsymbol{N}}
\newcommand{\hv}{\boldsymbol{h}}
\newcommand{\yv}{\boldsymbol{y}}
\newcommand{\RE}{\textrm{Re}}
\newcommand{\IM}{\textrm{Im}}
\newcommand{\rot}{\textrm{rot}}
\newcommand{\dv}{\boldsymbol{d}}
\newcommand{\grad}{\textrm{grad}}
\newcommand{\Tr}{\textrm{Tr}}
\newcommand{\ua}{\uparrow}
\newcommand{\da}{\downarrow}
\newcommand{\ct}{\textrm{const}}
\newcommand{\xv}{\boldsymbol{x}}
\newcommand{\mv}{\boldsymbol{m}}
\newcommand{\rv}{\boldsymbol{r}}
\newcommand{\kv}{\boldsymbol{k}}
\newcommand{\VE}{\boldsymbol{V}}
\newcommand{\sv}{\boldsymbol{s}}
\newcommand{\RV}{\boldsymbol{R}}
\newcommand{\pv}{\boldsymbol{p}}
\newcommand{\PV}{\boldsymbol{P}}
\newcommand{\EV}{\boldsymbol{E}}
\newcommand{\DV}{\boldsymbol{D}}
\newcommand{\BV}{\boldsymbol{B}}
\newcommand{\HV}{\boldsymbol{H}}
\newcommand{\MV}{\boldsymbol{M}}
\newcommand{\be}{\begin{equation}}
\newcommand{\ee}{\end{equation}}
\newcommand{\ba}{\begin{eqnarray}}
\newcommand{\ea}{\end{eqnarray}}
\newcommand{\bq}{\begin{eqnarray*}}
\newcommand{\eq}{\end{eqnarray*}}
\newcommand{\pa}{\partial}
\newcommand{\f}{\frac}
\newcommand{\FV}{\boldsymbol{F}}
\newcommand{\ve}{\boldsymbol{v}}
\newcommand{\AV}{\boldsymbol{A}}
\newcommand{\jv}{\boldsymbol{j}}
\newcommand{\LV}{\boldsymbol{L}}
\newcommand{\SV}{\boldsymbol{S}}
\newcommand{\av}{\boldsymbol{a}}
\newcommand{\qv}{\boldsymbol{q}}
\newcommand{\QV}{\boldsymbol{Q}}
\newcommand{\ev}{\boldsymbol{e}}
\newcommand{\uv}{\boldsymbol{u}}
\newcommand{\KV}{\boldsymbol{K}}
\newcommand{\ro}{\boldsymbol{\rho}}
\newcommand{\si}{\boldsymbol{\sigma}}
\newcommand{\thv}{\boldsymbol{\theta}}
\newcommand{\bv}{\boldsymbol{b}}
\newcommand{\JV}{\boldsymbol{J}}
\newcommand{\nv}{\boldsymbol{n}}
\newcommand{\lv}{\boldsymbol{l}}
\newcommand{\om}{\boldsymbol{\omega}}
\newcommand{\Om}{\boldsymbol{\Omega}}
\newcommand{\Piv}{\boldsymbol{\Pi}}
\newcommand{\UV}{\boldsymbol{U}}
\newcommand{\iv}{\boldsymbol{i}}
\newcommand{\nuv}{\boldsymbol{\nu}}
\newcommand{\muv}{\boldsymbol{\mu}}
\newcommand{\lm}{\boldsymbol{\lambda}}
\newcommand{\Lm}{\boldsymbol{\Lambda}}
\newcommand{\opsi}{\overline{\psi}}
\renewcommand{\tan}{\textrm{tg}}
\renewcommand{\cot}{\textrm{ctg}}
\renewcommand{\sinh}{\textrm{sh}}
\renewcommand{\cosh}{\textrm{ch}}
\renewcommand{\tanh}{\textrm{th}}
\renewcommand{\coth}{\textrm{cth}}

\begin{abstract}
An anisotropic fluid with variable energy density and negative pressure is proposed, both outside and inside stars. The gravitational field is constant everywhere in free space (if we neglect the local contributions) and its value is of the order of $g = 10^{~-8} cm/s^{2}$, in accordance with MOND model.

With $\rho,~ p \propto 1/r$, the acceleration is also constant inside stars but the value is different from one star to another and depends on their mass $M$ and radius $R$. In spite of the fact that the spacetime is of Rindler type and curved even far from a local mass, the active gravitational energy on the horizon is $-1/4g$, as for the flat Rindler space, excepting the negative sign. 
 
\end{abstract}

 \section{Introduction}
  A wide variety of current data supports the view that the matter content of the Universe consists of two basic components, namely dark matter (DM) and dark energy (DE) with ordinary matter playing a minor role. The nature and composition of DM and DE is not at all understood.
  
  As Mannheim \cite{PM1} has noticed, what is disturbing is the \textit{ad hoc} way in which DM is actually introduced. The DE problem is even more severe (its composition and structure is as mysterious as that of DM). The entire motivation for the existence of DM and DE is based on their validity at all distance scales of the standard Newton - Einstein gravitational theory \cite{PM1}.
  
  Whenever the theory is found to face observational difficulties on any particular distance scale, modifications are to be made to the stress tensor through the introduction of new, after the fact, gravitational sources so that agreement with observations is restored.
  
  While most of the researches have focused on the standard Newton - Einstein picture in Cosmology with a possible departure associated with a distance scale, Milgrom \cite{MM1} (see also \cite{MM2, LC, CH} suggested that the determining factor was not a distant scale but an acceleration one, with departure from the standard now appearing whenever accelerations of particles are less than some critical value $a_{0} = 10^{-8} cm/s^{2}$.
  
  From a different point of view, Mannheim \cite{PM2} obtained a 4-th order Einstein's equations starting with a conformally invariant Lagrangean - the Weyl tensor squared. These equations admit a static spherically symmetric vacuum solution that contains, besides the Schwarzschild term $const./r$, another term proportional to $r$ and, therefore, the metric is no longer asymptotically flat. In Mannheim's view, this linearly rising potential term shows that a local matter distribution can actually have a global effect at infinity and gravity theories become global. Using conformal invariance, he showed also that the vacuum solution has also a constant term which arises not from within a galaxy but comes from the global Hubble flow of the Universe itself, being related to its 3-space scalar curvature.
   
 Recently Grumiller \cite{DG} (see also \cite{RAS}) proposed a paradigm for gravitation at large distances. His metric generates a new Rindler acceleration term in a spherically symmetric situation, just as Mannheim stated before in his conformally invariant model of cosmology. The Rindler constant acceleration may depend on the scale of the system under consideration and becomes important at large distances from the source. 
  
 The Mannheim - Grumiller solution in four dimensions, without the cosmological constant, is given by 
 \begin{equation}
  ds^{2} = -(1- \frac{2M}{r} - 2gr) dt^{2} + (1- \frac{2M}{r} - 2gr)^{-1} dr^{2} + r^{2} d \Omega^{2}
 \label{1.1}
 \end{equation}
 where $M$ is the central mass, $g$ is the ''Rindler'' acceleration and $d \Omega^{2}$ stands for the metric on the unit 2 - sphere. To be a solution of the standard Einstein's equations one shows that a stress tensor is needed on the r.h.s. , namely
\begin{equation}
 T_{t}^{t} = - \rho = - \frac{g}{2 \pi r},~~~p_{r} = T_{r}^{r} = - \rho,~~~T^{\theta}_{\theta} = T^{\phi}_{\phi} = p_{\bot} = \frac{1}{2} p_{r}
 \label{1.2}
 \end{equation}
 where $\rho$ is the energy density of the anisotropic fluid, $p_{r}$ is the radial pressure and $p_{\bot}$ are the tangential pressures. To model the galactic rotation curves and Pioneer anomaly the condition $g > 0$ is to be imposed (notice that $g$ is scale dependent). 
 
 We analyse the spacetime (1.1) outside a spherical mass $M$, finding that it is endowed with two horizons when $M < 1/16g$ (the most ''obvious'' case). When $r >> \sqrt{M/g}$ (at large distances from the source), the horizon is located at $r = 1/2g$ and the geometry resembles Rindler's (it is not flat, of course, due to the singularity at $r = 0$ - the Kretschmann scalar is divergent there). 
  
 We show further in this paper that the active gravitational energy \cite{TP1} of the spacetime is negative because the strong energy condition (SEC) for the stress tensor is not satisfied. Its modulus equals the energy of the Rindler horizon obtained in \cite{HC1} (see also \cite{KM1}), when $M = 0$ (up to the radius $r$, we have $E = - gr^{2}$). We apply the model inside a relativistic star and reach the conclusion that the gravitational field $a$ is constant everywhere but depends from a star to another (the ''outside'' acceleration $g$ is negligible compared to $a$). Therefore, the spacetime resembles that one outside a planar domain wall.\\
 Throughout the paper we use the geometrical units $G = c = 1$.
 
 \section{Schwarzschild - Rindler spacetime}
 The effective energy - momentum tensor we need on the r.h.s. of Einstein's equations 
  \begin{equation}
  R_{~a}^{b} - \frac{1}{2} \delta_{a}^{b} R_{~a}^{b} = 8 \pi T_{~a}^{b} 
 \label{2.1}
 \end{equation}
 in order that the metric (1.1) to be a solution, is given by (1.2). We observe that the anisotropic fluid is comoving with the accelerated observer (the stress tensor is diagonal). 
 
 By means of Eqs. (1.1) and (1.2) we find that 
 \begin{equation}
 R_{~a}^{b} = diag(\frac{2g}{r}, \frac{2g}{r}, \frac{4g}{r}, \frac{4g}{r}),~~R_{~a}^{a} = \frac{12g}{r}, ~~R_{~~~~abcd}^{abcd} = \frac{16(3M^{2} + 2g^{2}r^{4})}{r^{6}}
\label{2.2}
\end{equation}
 While all the components of the Weyl tensor vanish when $M = 0$, the same property is not valid for the Riemann tensor. On the contrary, the components of the mixed stress tensor do not depend on $M$ but only on the Rindler acceleration $g$ (besides the radial distance $r$). In other words, $M$ plays no role as regards the dynamic parameters of the fluid.
 
 Let us write down the general expression for the energy - momentum tensor for an anisotropic fluid \cite{HC2, KM2}
 \begin{equation}
T_{a}^{b} = (\rho + p_{\bot})u_{a} u^{b} + p_{\bot} \delta_{a}^{b}+ (p_{r} - p_{\bot}) s_{a} s^{b}.
\label{2.3}
\end{equation}
 In (2.3), $u^{a}$ is the timelike velocity vector of the fluid and $s^{a}$ is spacelike , on the direction of anisotropy, with $s^{a} u_{a} = 0$. The stress tensors (2.3) and (1.2) coincide provided
 \begin{equation}
 u^{a} = (\frac{1}{\sqrt{1- \frac{2M}{r} - 2gr}}, 0, 0, 0)~,~~~s^{a} = (0, \sqrt{1- \frac{2M}{r} - 2gr}, 0, 0) 
\label{2.4}
\end{equation}
Let us find now the location of the horizons in the metric (1.1), obtained from
 \begin{equation}
 1- \frac{2M}{r} - 2gr = 0.
\label{2.5}
\end{equation}
We have two horizons located at
 \begin{equation}
 r_{\pm} = \frac{1 \pm \sqrt{1 - 16 g M}}{4g},
\label{2.6}
\end{equation}
when $M < 1/16g$, the most encountered situation. We shall adopt for $g$ the value $g = 10^{-8} cm/s^{2}$ for the horizon located at the distance $1/2g$ to correspond to the radius of the visible universe, $R_{U} \approx 10^{28} cm$. This will be in accordance with MOND acceleration and the Pioneer anomaly. 

 By means of the expression from (1.2) we can show that the above value of $g$ is valid everywhere in our Universe. For example, the mass $m(r)$ inside a sphere of radius $r$ (with $2gr >> 2M/r$, or $r >> \sqrt{M/g} \equiv r_{crit}$, which corresponds to $r \approx 5.10^{16} cm$ if $M$ is the mass of the Sun) will be \cite{SW}
\begin{equation}
 m(r) = \int{\rho(r) \sqrt{- \gamma} dV} = \int{4 \pi r^{2} \rho(r) dr = gr^{2}}, 
\label{2.7}
\end{equation}
taken from $0$ to $r$ (the term $2M/r$ in the metric has been neglected and $\gamma$ stands for the determinant of the metric) . Therefore, the gravitational field at $r$ is $m(r)/r^{2} = g = const$. The result can explain why the ''critical acceleration'' $10^{-8} cm/s^{2}$ in MOND theory coincides with the acceleration $1/R_{U}$ at the universe horizon \cite{LS}. The above value of $g$ seems to be universal if the present model proves to be correct.

 The case $M << 1/16g$ leads to $r_{-} \approx 2M$ and $r_{+} \approx 1/2g$. At $M = 1/16g$, we have $r_{-} = r_{+} = 1/4g = r_{H}$. In terms of $r_{H}$, (2.7) can be written as  
 \begin{equation}
 r_{\pm} = r_{H} (1 \pm \sqrt{1 - \frac{4M}{r_{H}}}).
\label{2.8}
\end{equation}
There is no horizon for $M > 1/16g$. It is worth noting that $g_{tt}$ in (1.1) is positive only for $r \in(r_{-}, r_{+})$; otherwise $r$ becomes timelike and the geometry (1.1) will be nonstationary. In addition, $r_{-}$ is located deeply inside the star, where the metric (1.1) is not valid. We also observe that, on the outer horizon, the energy density of the anisotropic fluid is given by 
\begin{equation}
  \rho_{+} = \frac{g}{2 \pi r_{+}} = \frac{2g^{2}}{\pi (1 + \sqrt{1 - 16gM})}.
\label{2.9}
\end{equation}
The condition $M << 1/16g$ leads to $\rho = g^{2}/\pi$, namely $\rho \propto g^{2}$. A similar dependence of the gravitational energy density on acceleration has been obtained by Padmanabhan \cite{TP2} (see also \cite{HC3}).

Let us compute now the active gravitational energy of the anisotropic fluid. We have \cite{TP1}
\begin{equation}
E = 2 \int(T_{ab} - \frac{1}{2} g_{ab}T)u^{a} u^{b} N\sqrt{\gamma} d^{3}x ,
\label{2.10}
\end{equation}
where $u^{a}$ is given by (2.4), $N = \sqrt{-g_{tt}}$ is the lapse function and $\gamma$ is the determinant of the spatial 3 - metric. One obtains
\begin{equation}
E = -g (r_{+}^{2} - R^{2}) ,
\label{2.11}
\end{equation}
where the integration has been performed for $r \in[R, r_{+}]$, $R$ being the star radius. Eq. (2.11) yields
\begin{equation}
E = - \frac{1}{8g} (1 +  \sqrt{1 - 16GM}) + M + gR^{2}.
\label{2.12}
\end{equation}
The fact that $E < 0$ (for any reasonable $M$ and $R$) is rooted from the negative pressure contribution (the energy - momentum (1.2) does not obey the strong energy condition).

If we remove the mass $M$, the gravitational energy becomes $E = -1/4g$, a value already obtained in a previous paper \cite{HC1} using different arguments (see also \cite{KM1}) for the Rindler flat metric. However, in the current situation the geometry is not Minkowskian (the Riemann tensor is nonzero even when $M = 0$). Numerically, one obtains $E = -1/4g \approx - 10^{77} erg$, namely $E/c^{2} \approx - 10^{56} g$. This value is equal to the mass of the universe, excepting the negative sign. It supports the old suggestion that the total energy of the universe is vanishing.

\section{Inside a relativistic star}
 Let us see how the previous model works inside a spherically symmetric relativistic star of mass $M$ and radius $R$. At the scale of a star, the term $-2gr$ may be neglected (it counts only at galactic scale).
 
 Take an observer located at some distance $r$ from the star center. The gravitational field there depends only on the mass $m(r)$ up to the radius $r$. Therefore, one may remove the mass ''above'' the observer and the metric becomes Schwarzschild's, with $g_{tt} = -1 + 2 m(r)/r$. We conjecture the form of the stress tensor (1.3) is also valid inside the star. Therefore, $m(r)$ is given by \cite{SW}
\begin{equation}
m(r) = \int{4 \pi r^{2} \rho(r) dr} = a r^{2} ,
\label{3.1}
\end{equation}
 where the integral is taken from $0$ to $r$. We replaced $g$ (valid outside the star) with the constant acceleration $a$ so that $\rho = a/2 \pi r$ ( $a$ is specific to any particular star ; for a star of mass $M$ and radius $R$, $a$ is given by its surface value, namely $a = M/R^{2})$. Taking $m(r)$ from (3.1), one obtains $g_{tt} = -1 + 2ar$. Consequently, the inner metric appears as 
\begin{equation}
 ds^{2} = -(1 - 2ar) dt^{2} + (1 - 2ar)^{-1} dr^{2} + r^{2} d \Omega^{2},
\label{3.2}
\end{equation}
 with a horizon at $r = 1/2a$. 
 
 Let us write the energy density under the form
 \begin{equation}
 \rho = - p_{r} = \frac{2 \sigma}{r} ,
\label{3.3}
\end{equation}
where $\sigma$ is a positive constant. Its meaning outcomes by writing the expression of the mass $m(r)$ ($r < R$)
 \begin{equation}
 m(r) = \int{4 \pi r^{2} \rho(r) dr} = 4 \pi \sigma r^{2}.
\label{3.4}
\end{equation}
In other words, $\sigma$ may represent the surface tension on $r = const.$ (Eq. (3.3) remind us the Young - Laplace formula). One might consider that all the mass up to the radius $r$ is concentrated on a thin shell of radius $r$ (from the gravitational viewpoint, Birkhoff's theorem may be used as an evidence of the equivalence of the two physical situations ; the Holographic Principle leads to a similar conclusion).

So far as $r \rho(r)$ is constant, we may find $\sigma$ from
 \begin{equation}
 r \rho(r) = R \rho(R) = 2 \sigma .
\label{3.5}
\end{equation}
 Hence, $\sigma = M/4 \pi R^{2}$. With, say, $\rho(R) = 1.4~ g/cm^{3}$ and $R = 7.10^{10} cm$, we get $\sigma \approx 5.10^{10} g/cm^{2}$ or the surface energy density $\sigma c^{2} \approx 10^{31} erg/cm^{2}$. The acceleration at distance $r$ from the center is $a = m(r)/r^{2} = 4 \pi \sigma$ (by using, for instance, Gauss' theorem). The fact that, according to our model, the gravitational field is constant inside the star brings to our attention the acceleration near a planar wall \cite{IS, PJ, ED, RGS, MJ}, in spite of the spherical symmetry. A clue to the explanation of this similarity would be given by Ipser and Sikivie \cite{IS} who remarked that ''In the Minkowski coordinates...the planar wall is not a plane at all but rather an accelerated sphere...''.
 
 When the star surface is viewed as an interface of phase transition, the surface tension may be obtained \cite{CG, SM} from $[p_{r}] = 2 \sigma /r$, where $[p_{r}] = p_{r}^{out} - p_{r}^{in}$ is the jump in the radial pressure (evaluated on the surface), with $p_{r}^{out} = 0$. We again obtain $\rho = 2 \sigma/r$, the surface tension compensating the pressure gradient. Since the anisotropic ''force'' $(2/r)(p_{\bot} - p_{r}) = \rho/r$ is positive, it will be directed outward \cite{FR}, having a repulsive character. This is a consequence of the fact that SEC is not fulfilled, as is obvious for dark energy which has a repulsive nature. 

 A problem of the model is related to the fact that $\rho \propto 1/r$, a property that seems not to be valid experimentally (at cosmological scale). We must keep in mind that inside a star we have a localized center, contrary to the cosmological situation. Moreover, there is a true singularity at $r = 0$, like in the Schwarzschild case (the Kretschmann scalar is infinite there). As far as the ''Rindler'' horizon is concerned, its location at $r = 1/2a$ is beyond any actual radius of a star. With $a = M/R^{2}$ we obtain $1/2a = R^{2}/R_{g}$, where $R_{g} = 2M$ is the gravitational radius of the star (in the case the star becomes a black hole, the ''Rindler'' horizon and the black hole horizon overlap). It is worth to note that the actual radius is the geometrical mean between the Schwarzschild radius and the Rindler horizon radius of an inner observer ; it is interesting that the horizon is in either case located in a region where the metric is no longer valid : for the inner observer it is outside and for the exterior one the other horizon is located inside. Taking the Sun as an example, with $M = 2.10^{33} g$ and $R = 7.10^{10} cm$, we have $1/2a = 10^{16} cm$, much more than its actual radius. Therefore, the horizon of the geometry (1.1) is obviously located in a region where the metric is not  valid (the star terminates at $R << 1/2a)$.
 
 We wish to mention that Mannheim \cite{PM1} used also an anisotropic fluid in a model for a relativistic star. From the conservation of $T_{a}^{b}$ he obtained
 \begin{equation}
 \frac{dp_{r}}{dr} + \frac{\rho + p_{r}}{2B} \frac{dB}{dr} + \frac{2}{r} (p_{r} - p_{\bot}) = 0 ,
\label{3.6}
\end{equation}
where $B = - g_{tt}$ and the last term is vanishing in the standard model \cite{SW}. Mannheim observed that, in passing from flat to curved space, it is not mandatory to preserve the perfect fluid model with $p_{r} = p_{\bot}$ because the transition is not kinematic but a dynamic one. It is easy to check that his eq. (49) is obeyed when $p_{r} = - a/2 \pi r$ (with his $q = p_{\bot} = p_{r}/2$) in the geometry (1.1).

\section{The junction conditions}
Leu us now study the matching conditions at the star surface $S$. We have (3.2) as the metric inside the star and the Schwarzschild geometry 
\begin{equation}
 ds^{2} = -(1 - \frac{2M}{r}) dt^{2} + (1 - \frac{2M}{r})^{-1} dr^{2} + r^{2} d \Omega^{2},
\label{4.1}
\end{equation}
outside it. Because $a = M/R^{2}$, it is clear that the first fundamental forms (the metric coefficients) are equal on the surface. We shall further use the Gauss - Codazzi formalism \cite{IS} (see also \cite{CL}) to analyze the jump of the extrinsic curvature (the second fundamental form) when the surface (viewed as a thin shell) is crossed.

The unit spacelike normals of $S$ are given by 
\begin{equation}
n_{-}^{a} = (0, \sqrt{1 - 2ar}, 0, 0), ~~~n_{+}^{a} = (0, \sqrt{1 - \frac{2M}{r}}, 0, 0)
\label{4.2}
\end{equation}
where $(-/+)$ means (inside/outside). The metric induced on $S$ will be $h_{ab} = g_{ab} - n_{a} n_{b}$. For the extrinsic curvature of the surface we have 
\begin{equation}
K_{ab} = h_{a}^{c} \nabla_{c} n_{b}
\label{4.3}
\end{equation}
The discontinuity of the extrinsic curvature $[K_{ab}] = K_{ab}^{+} - K_{ab}^{-}$ is related to the stress tensor $S_{ab}$ on the hypersurface $S$ by the Lanczos equation
\begin{equation}
[K_{ab}] - h_{ab} [K] = -8 \pi S_{ab}
\label{4.4}
\end{equation}
where $K$ is the trace of $K_{ab}$. 

We take $S_{ab}$ of the form \cite{CL} 
\begin{equation}
S_{ab} = (\Sigma + p_{s}) u_{a} u_{b} + p_{s} h_{ab}
\label{4.5}
\end{equation}
where $\Sigma$ stands for the surface energy density, $u_{a}$ is a timelike unit vector orthogonal to $n^{a}$ and $p_{s}$ is the surface pressure. Using Eq. (4.2) - (4.5) one finds that 
\begin{equation}
\begin{split}
K_{t,-}^{t} = \frac{-a}{\sqrt{1 - 2aR}},~K_{t,+}^{t} = \frac{M}{R^{2} \sqrt{1 - \frac{2M}{R}}},~ K_{\theta,-}^{\theta} = K_{\phi,-}^{\phi} = \frac{1}{R} \sqrt{1 - 2aR} \\
K_{\theta,+}^{\theta} = K_{\phi,+}^{\phi} = \frac{1}{R} \sqrt{1 - \frac{2M}{R}},~K_{-} = \frac{2 - 5aR}{R \sqrt{1 - 2aR}},~K_{+} = \frac{2R - 3M}{R^{2} \sqrt{1 - \frac{2M}{R}}}
\label{4.6}
\end{split}
\end{equation}
 Let us observe that we have no a jump of the angular components of the extrinsic curvature. However
 \begin{equation}
 [K_{t}^{t}] = [K] = \frac{2M}{R^{2} \sqrt{1 - \frac{2M}{R}}}
\label{4.7}
\end{equation}
Therefore, (4.4) yields $S_{t}^{t} = 0$, which leads to $\Sigma = 0$. On the contrary, the $\theta \theta$ - components of Eq. (4.4) will give $- h_{\theta \theta} [K] = - 8 \pi S_{\theta \theta}$ . Hence $p_{s} = M/4 \pi R^{2}\sqrt{1 - 2M/R}$. We observe that $p_{s}$ is not equal to $\sigma$ from (3.5). The reason comes from the classical character of the Young - Laplace equation (3.3). Its relativistic counterpart \cite{CDR} has $\sigma K$ instead of $2 \sigma/r$ on the r.h.s., where $K = K_{-}$ from (4.6). Taking $2M/r << 1$, $K_{-} = 2/R$ and the classical expression is restored.

It is worth to note that we did not impose $\Sigma = 0$ as Ghezzi \cite{CG} did with his surface energy density $\eta$ ; instead, the vanishing $\Sigma$ was a consequence of our equations.

\section{Conclusions} 
 A nonuniform model for the geometry in the interior of a relativistic star is developed, on the basis of an anisotropic fluid stress tensor with negative radial and tangential pressures. The energy density falls off as $1/r$ when we move away from the center but the gravitational field $a$ is constant everywhere, as for a planar wall, being specific to any star. We have a true singularity at $r = 0$ and a horizon at $r = 1/2a$ which normally is located outside the star.\\ 
 Far from the mass $M$ the dominant term in the metric is $-2gr$, where the constant ''cosmological'' acceleration $g$ is $\approx 10^{-8} cm/s^{2}$, the ''critical'' value appearing in MOND model. It is also of the order of the anomalous  acceleration encountered in the Pioneer travel and equals the surface gravity of the horizon of the Universe. \\
 We assumed also that the value $10^{-8} cm/s^{2}$ represents a universal acceleration because of its constancy at any scale.\\
 
 \textbf{Acknowledgements} \\
 I would like to thank T. Padmanabhan for helpful comments and suggestions.

\end{document}